\documentclass[cernpreprint, atlasdraft=false, texlive=2023, USenglish, orcidlogo, cmslogo=true, texmf, orcidlogo]{atlasdoc}
 
\usepackage{atlaspackage}
\usepackage{atlasbiblatex}
 
\usepackage{atlascontribute}
 
\usepackage{atlasphysics}

\graphicspath{{logos/}{figures/}}
 
\usepackage{ANA-HIGG-2022-22-PAPER-defs}

\AtlasTitle{\vspace{0.5cm}
Evidence for the Higgs boson decay to a $Z$ boson and a Photon at the LHC}
 
\AtlasVersion{2.0}
 
\AtlasAbstract{
The first evidence for the Higgs boson decay to a $Z$ boson and a photon is presented, with a statistical significance of 3.4 standard deviations. The result is derived from a combined analysis of the searches performed by the ATLAS and CMS Collaborations with proton--proton collision data sets collected at the CERN Large Hadron Collider (LHC) from 2015 to 2018. These correspond to integrated luminosities of around 140 fb$^{-1}$ for each experiment, at a center-of-mass energy of 13 TeV. The measured signal yield is $2.2\pm0.7$ times the standard model prediction, and agrees with the theoretical expectation within 1.9 standard deviations.
}

\AtlasRefCode{ANA-HIGG-2022-22}
\PreprintIdNumber{CERN-EP-2023-114}
\AtlasJournalRef{Phys. Rev. Lett. 132 (2024) 021803}
\AtlasDOI{10.1103/PhysRevLett.132.021803}

\hypersetup{pdftitle={ATLAS-CMS document},pdfauthor={The ATLAS and CMS Collaborations}}
 
\begin{document}
 
\maketitle

% The next lines are included from the .//sections/introduction.tex input file
 
Since the discovery of the Higgs boson~\cite{HIGG-2012-27, CMS-HIG-12-028, CMS-HIG-12-036} by the ATLAS~\cite{PERF-2007-01} and CMS~\cite{CMS-CMS-00-001} Collaborations in 2012, a detailed program of measurements~\cite{HIGG-2015-07,HIGG-2021-23,CMS-HIG-22-001} has confirmed its couplings and other properties to be mostly consistent with those predicted by the Standard Model (SM).
However, there are several rare Higgs boson decay channels, including \HZg~\cite{HZY_origin_1,HZY_origin_2,HZY_origin_3}, that have not been observed.
These channels provide probes for possible contributions arising from physics beyond the SM (BSM physics). During LHC \RunTwo (2015--2018), large data samples of proton--proton collisions at $\sqrt{s} = 13\,\TeV$ were collected by the two experiments, improving the sensitivity to such decays.
 
\begin{figure}[htbp]
\centering
\subfloat[]{\includegraphics[width=0.7\textwidth]{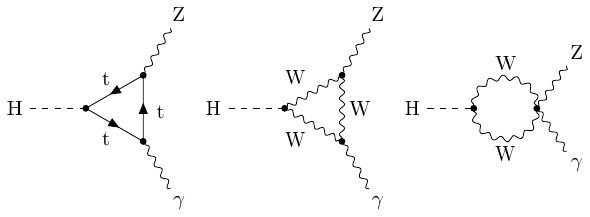}}
\caption{Examples of Feynman diagrams for \HZg decay.
}
\label{fig:diagrams}
\end{figure}

In the SM, the \HZg decay is expected to have a relatively small branching fraction of $(1.5\pm 0.1) \times 10^{-3}$ for a Higgs boson mass ($m_{H}$) close to 125\,\GeV~\cite{Djouadi:1997yw,deFlorian:2016spz}.
As the \HZg decay occurs via loop diagrams, with examples given in Figure~\ref{fig:diagrams}, it is sensitive to modifications in several BSM scenarios that would cause the branching fraction to be enhanced compared with the SM value. Examples include models where the Higgs boson
is a composite state~\cite{azatov}, a pseudo Nambu--Goldstone boson~\cite{HVVHZyRatio}, or a neutral scalar originating from a different source~\cite{low, low2}.
Branching fractions deviating from the SM value are also expected for models with additional colorless charged scalars, leptons or vector bosons that couple to the Higgs boson, because of their contributions via loop corrections~\cite{carena, extensions, Chen:2013vi}.
 
This Letter reports the first evidence for \HZg decay, obtained from a combination of ATLAS~\cite{HIGG-2018-42} and CMS~\cite{CMS-HIG-19-014} searches for this channel.
The analyses are based on the \RunTwo data sets collected by the ATLAS and CMS experiments, corresponding to integrated luminosities of $139$ and $138\,\mathrm{fb}^{-1}$, respectively, at a center-of-mass energy of $13\,\TeV$.
Previous $\HZg$ searches by the ATLAS and CMS Collaborations used the data sets collected at $\sqrt{s} = 7$ and $8\,\TeV$, and partial data sets collected at $\sqrt{s} = 13\,\TeV$~\cite{HIGG-2013-05,CMS-HIG-13-006,CMS-HIG-17-007}.
 
The ATLAS detector~\cite{PERF-2007-01} is a multipurpose particle detector with cylindrical geometry. It consists of an inner tracking detector surrounded by a thin superconducting solenoid providing a $2\,\unit{T}$ axial magnetic field, electromagnetic and hadronic sampling calorimeters, and a muon spectrometer with three toroidal superconducting magnets, providing nearly $4\pi$ coverage in solid angle.
The CMS apparatus~\cite{CMS-CMS-00-001} is a nearly hermetic, multipurpose detector. Contained within a $3.8\,\unit{T}$ superconducting solenoid are an all-silicon inner tracker, a crystal electromagnetic calorimeter, and a brass--scintillator hadron calorimeter. Gas-ionization muon detectors are embedded in the flux-return yoke outside the solenoid.

The ATLAS and CMS $\HZg$ analyses share many features.
In both, the $Z$ boson is reconstructed through its decays into electron or muon pairs ($Z\rightarrow\ell^{+}\ell^{-}$, $\ell$ = $e$ or $\mu$), requiring a dilepton mass above $50\,\GeV$. The leptons provide a clean signature and ensure a high trigger efficiency and good invariant mass resolution for the final-state products of the Higgs boson decay.
The photon candidate is reconstructed from energy clusters in the electromagnetic calorimeters. It must satisfy identification criteria and be isolated from other event activity.
The dominant backgrounds arise from Drell--Yan production in association with a photon or a jet misidentified as a photon. In both analyses, the production of the SM Higgs boson signal is modeled with the \POWHEGBOX[v2] Monte Carlo event generator~\cite{Nason:2009ai,Alioli:2010xd,Nason:2004rx,Frixione:2007vw,Hartanto:2015uka,powhegGGH}.
 
After the reconstruction and selection of \HZg candidate events, the signal is identified as a narrow resonant peak at $m_{H}$ in the $Z\gamma$ invariant mass ($m_{Z\gamma}$) distribution, calculated as the invariant mass of the $\ell^+\ell^-\gamma$ system.
The $m_{Z\gamma}$ resolution is improved by dedicated final-state radiation corrections to the momenta of muons with nearby photons, and via kinematic fits for the dilepton mass of the $Z$ boson candidate using a Breit--Wigner line shape to model the $Z$ boson resonance, convolved with a Gaussian response function for the leptons~\cite{HIGG-2013-21,CMS-HIG-16-041}.
The resulting $m_{Z\gamma}$ resolution is $1.4$--$2\,\GeV$, depending on the final state and event topology.

% End of text imported from the .//sections/introduction.tex input file

% The next lines are included from the .//sections/experiments_simple.tex input file
To enhance the sensitivity, both analyses assign events to categories with different signal-to-background ratios by exploiting the kinematic features of different Higgs boson production modes.
The ATLAS analysis assigns each event to one of six categories, including a category targeting the vector-boson fusion (VBF) topology, which requires the presence of at least two jets and a selection on the output score of a dedicated boosted decision tree (BDT).
The remaining five categories target other Higgs boson production modes, which are defined with different lepton flavors and kinematic properties of the momentum of the $Z\gamma$ system transverse to the beam direction~\cite{VESTERINEN2009432}.
 
The CMS analysis assigns each event to one of eight categories, including a category with additional leptons targeting the production of Higgs bosons associated with either a weak vector boson or a top quark pair, and three categories defined by the output score of a dedicated BDT targeting the VBF topology.
The other four categories are defined by the output score of another BDT exploiting the differences between the kinematic properties of \HZg signal events and background events.
 
Simultaneous signal-plus-background fits across the analysis categories are performed to the $Z\gamma$ invariant mass distribution, with analytic signal and background functions. The signal models are from Crystal Ball~\cite{dscb1} and Gaussian functions, and the background models are based on exponential functions, power law functions, Laurent series, and Bernstein polynomials. Experimental and theoretical uncertainties affecting the expected number of signal events, the shape of the $Z\gamma$ invariant mass distribution from the \HZg signal process, and the background models are considered as constrained nuisance parameters.
The Higgs boson production cross-sections and \HZg branching fraction used to normalize the signal are common to both experiments. In both analyses the parameters of the analytic background functions are determined from the data.
Both the  ATLAS and CMS analyses measure the signal strength ($\mu$),
defined as the ratio of the Higgs boson production cross-section times \HZg branching fraction to the SM prediction.

% End of text imported from the .//sections/experiments_simple.tex input file

% The next lines are included from the .//sections/procedure.tex input file
 
The statistical treatment of the data is based on the standard LHC data modeling and handling toolkits: \RooFit~\cite{RooFit}, \RooStats~\cite{RooStats}, and {\scshape HistFactory}\xspace~\cite{CERN-OPEN-2012-016}.
The confidence intervals of the signal strength are determined via the profile-likelihood-ratio test statistic~\cite{Cowan:2010js}.
The likelihood function used to define the test statistic is the product of the likelihood functions of the ATLAS and CMS analyses, adapted to have common constraint terms for the nuisance parameters representing the correlated uncertainties.
The main differences between the likelihood functions of the two analyses are the background models. In each category of the ATLAS analysis, the chosen background model is the one that minimizes the presence of \enquote{spurious} signal, i.e.\
the extracted signal yields in signal-plus-background fits to background-only templates of $m_{Z\gamma}$~\cite{HIGG-2012-27}. The spurious-signal yield is introduced as an additional nuisance parameter in the likelihood function, which modifies the expected signal yield. In the CMS analysis, a discrete profiling method~\cite{dauncey2015handling} is used to determine the background model directly in the fit to the data. For each category, this method introduces an additional discrete nuisance parameter that selects the best background model among a large set of alternative models. The discrete nuisance parameter is profiled in the final fit.
 
The experimental uncertainties from the ATLAS and CMS analyses are considered uncorrelated. While some components of the experimental systematic uncertainties could be correlated due to the similar simulation software and calibration techniques, these are expected to be much smaller than the uncorrelated components.
Among the theoretical uncertainty sources, the one associated with missing higher orders in the perturbative calculations of the gluon--gluon fusion cross-section (renormalization and factorization scale uncertainties)~\cite{Stewart:2011cf,Gangal:2013nxa,Grazzini:2013mca}, and the ones in the \HZg branching fraction prediction~\cite{deFlorian:2016spz} are correlated.
In the CMS analysis, a small number of modifications are made to facilitate the combination. In particular, for consistency, the scale and branching fraction uncertainties are re-evaluated. These changes have a minor impact on the CMS result.
In the ATLAS analysis, the decomposition of the scale uncertainties in terms of independent sources is modified, which has a negligible impact on the observed signal strength.
The uncertainties associated with missing higher orders in the calculations for the other Higgs boson production modes, with the choice of parton distribution functions (PDF)~\cite{PDF_cms_1, PDF_cms_2}, with the value of the strong-force coupling constant (\alphas), and with the modeling of the underlying event and parton shower are not correlated because of their different implementations in the two analyses.
However, approximate correlation strategies were investigated for the integrated luminosity~\cite{ATLAS-CONF-2019-021,LUCID2,CMS-LUM-17-003,CMS-PAS-LUM-17-004,CMS-PAS-LUM-18-002}, scale and PDF uncertainties; they are found to have a negligible impact and are not adopted in the following results.
One difference between the ATLAS and CMS analyses is the assumed value of $m_{H}$, taken to be 125.09\,\GeV~\cite{HIGG-2014-14} in the former, and 125.38\,\GeV~\cite{CMS-HIG-19-004} in the latter.
The results of the combination are determined for both $m_{H}$ values, and the different mass assumptions have a negligible impact within the precision reported in this Letter.

% End of text imported from the .//sections/procedure.tex input file

% The next lines are included from the .//sections/results.tex input file
 
\begin{figure}[htbp]
\centering
\includegraphics[width=0.6\textwidth]{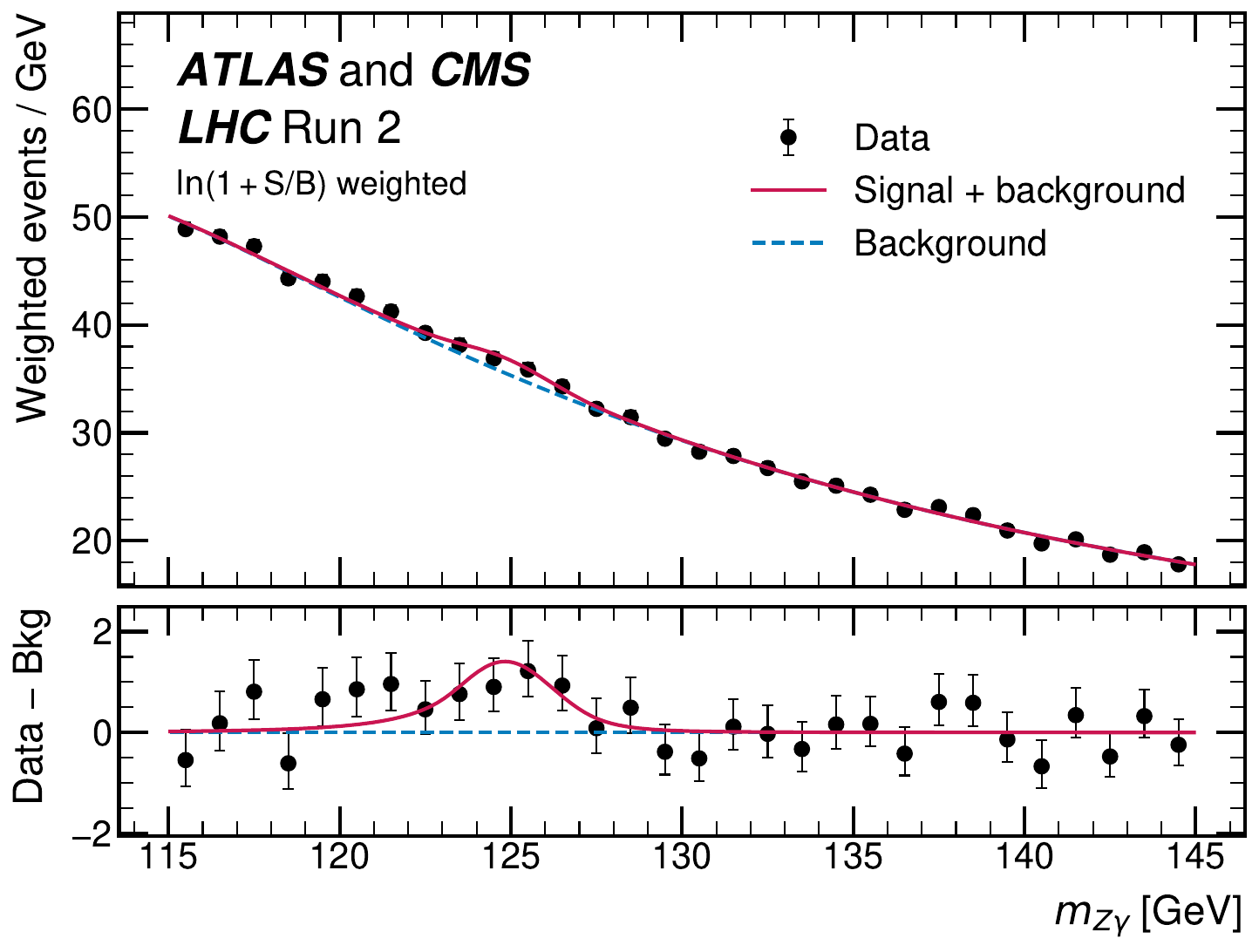}
\caption{
The $Z\gamma$ invariant mass distribution. Events from all categories in the ATLAS and CMS analyses are shown.
As different ranges in $m_{Z\gamma}$ are used in the two analyses, only the common subrange is visualized here.
The data (points with error bars) in each category are weighted by $\ln(1 + S/B)$, where $S$ and $B$ are the observed signal and background yields in that category, in the $120$--$130\,\GeV$ interval.
The $S$ and $B$ values are derived from the fit to data.
The error bars are invisible because of their small values. The fitted signal-plus-background (background) probability density functions (pdfs) in each category are also weighted in the same way and summed, and represented  by a red solid (blue dashed) line.
The lower panel shows the background-subtracted results with the same data and pdfs.
}
\label{fig:mass}
\end{figure}

\begin{figure}[htbp]
\centering
\includegraphics[width=0.6\textwidth]{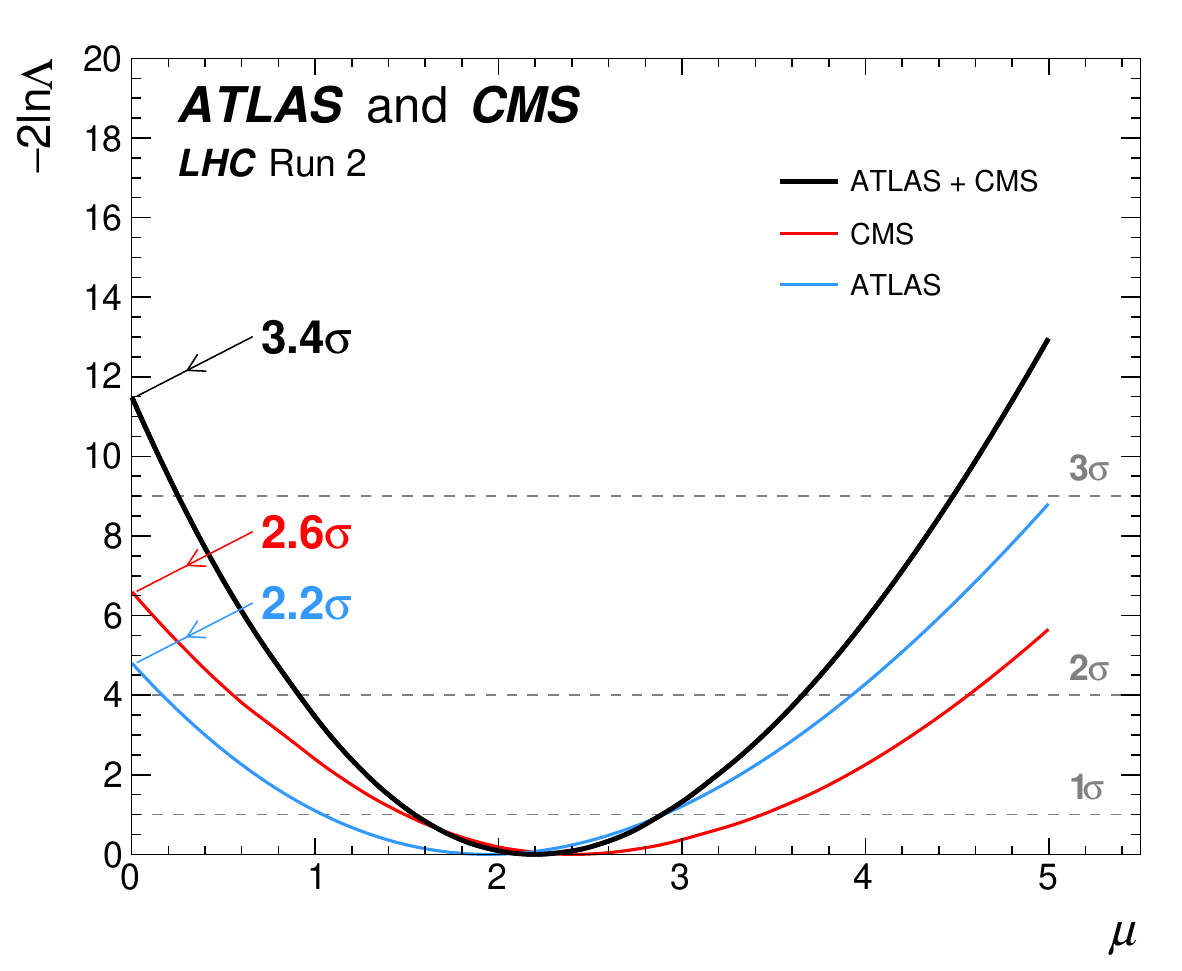}
\caption{
The negative profile log-likelihood test statistic, where $\Lambda$ represents the likelihood ratio, as a function of the signal strength $\mu$ derived from the ATLAS data (blue line), the CMS data (red line), and the combined result (black line).
The different Higgs boson masses assumed by ATLAS and CMS have a negligible impact on the results.
}
\label{fig:Comb}
\end{figure}

The $Z\gamma$ invariant mass distribution observed in data is shown in Figure~\ref{fig:mass}. To demonstrate the sensitivity of this likelihood analysis, the events in each category are weighted by $\ln(1 + S/B)$, where $S$ and $B$ are the observed signal and background yields in that category in the range $120 < m_{Z\gamma} < 130\,\GeV$, as determined by the minimization of the test statistic. The negative log-likelihood ratio as a function of the signal strength is shown in Figure~\ref{fig:Comb}.
The observed (expected) signal strength at the 68\% confidence level is $\mu = 2.0^{+1.0}_{-0.9}$ ($1.0\pm0.9$) for the ATLAS analysis, $\mu = 2.4^{+1.0}_{-0.9}$ ($1.0^{+1.0}_{-0.9}$) for the CMS analysis, and $\mu = 2.2\pm 0.6~\text{(stat.)} ^{+0.3}_{-0.2}~\text{(syst.)} = 2.2\pm0.7$ ($1.0\pm0.6~\text{(stat.)}\pm0.2~\text{(syst.)}=1.0\pm0.6$) for their combination.
Expressed in standard deviations, $\sigma$, the observed (expected) local significance, with respect to the $\mu=0$ hypothesis of no \HZg signal, is 2.2$\sigma$ (1.2$\sigma$) for the ATLAS analysis, 2.6$\sigma$ (1.1$\sigma$) for the CMS analysis, and 3.4$\sigma$ (1.6$\sigma$) for their combination.
The uncertainties in the \HZg branching fraction and the background modeling are the largest systematic uncertainties.
Assuming SM Higgs boson production cross-sections, the measured branching fraction for \HZg decay is $(3.4\pm1.1)\times10^{-3}$.
In contrast to the signal strength measurement, the uncertainty in the SM branching fraction prediction is not included in this fit.
The uncertainties in the results are dominated by the statistical fluctuations of data.
 
The combined result is compatible with the measured signal strengths from individual categories with a $p$-value greater than 12\%. The $p$-value for  compatibility with the SM hypothesis ($\mu=1$) is about 6\%,
and the observed local significance with respect to the SM is 1.9$\sigma$.
The goodness-of-fit of the model to the data is evaluated with a likelihood-ratio test~\cite{Baker:1983tu}, and has a $p$-value greater than 90\%.
Tabulated results are provided in the HEPData record for this analysis~\cite{hepdata}.
 
% End of text imported from the .//sections/results.tex input file

% The next lines are included from the .//sections/conclusion.tex input file
 
In summary, a combined analysis of ATLAS and CMS searches for the Higgs boson decay to a $Z$ boson and a photon, where the $Z$ boson decays into an electron or muon pair, is presented.
The results are based on the $13\,\TeV$ proton--proton collision data recorded by the ATLAS and CMS experiments at the CERN LHC, amounting to integrated luminosities of $139\,\mathrm{fb}^{-1}$ and $138\,\mathrm{fb}^{-1}$ respectively.
Evidence for \HZg decay is established, with an observed significance of 3.4 standard deviations. The observed signal yield is $2.2 \pm 0.7$ times the SM prediction. The measured \HZg branching fraction is $(3.4\pm1.1)\times10^{-3}$. The result agrees with the SM prediction within 1.9 standard deviations.

\newpage

% End of text imported from the .//sections/conclusion.tex input file

\section*{Acknowledgements}

% The next lines are included from the .//acknowledgements/Acknowledgements.tex input file

ATLAS thanks CERN for the very successful operation of the LHC, as well as the
support staff from our institutions without whom ATLAS could not be
operated efficiently.
 
ATLAS acknowledges the support of
ANPCyT, Argentina;
YerPhI, Armenia;
ARC, Australia;
BMWFW and FWF, Austria;
ANAS, Azerbaijan;
CNPq and FAPESP, Brazil;
NSERC, NRC and CFI, Canada;
CERN;
ANID, Chile;
CAS, MOST and NSFC, China;
Minciencias, Colombia;
MEYS CR, Czech Republic;
DNRF and DNSRC, Denmark;
IN2P3-CNRS and CEA-DRF/IRFU, France;
SRNSFG, Georgia;
BMBF, HGF and MPG, Germany;
GSRI, Greece;
RGC and Hong Kong SAR, China;
ISF and Benoziyo Center, Israel;
INFN, Italy;
MEXT and JSPS, Japan;
CNRST, Morocco;
NWO, Netherlands;
RCN, Norway;
MEiN, Poland;
FCT, Portugal;
MNE/IFA, Romania;
MESTD, Serbia;
MSSR, Slovakia;
ARRS and MIZ\v{S}, Slovenia;
DSI/NRF, South Africa;
MICINN, Spain;
SRC and Wallenberg Foundation, Sweden;
SERI, SNSF and Cantons of Bern and Geneva, Switzerland;
MOST, Taipei;
TENMAK, T\"urkiye;
STFC, United Kingdom;
DOE and NSF, United States of America.
In addition, individual groups and members have received support from
BCKDF, CANARIE, CRC and DRAC, Canada;
PRIMUS 21/SCI/017 and UNCE SCI/013, Czech Republic;
COST, ERC, ERDF, Horizon 2020, ICSC-NextGenerationEU and Marie Sk{\l}odowska-Curie Actions, European Union;
Investissements d'Avenir Labex, Investissements d'Avenir Idex and ANR, France;
DFG and AvH Foundation, Germany;
Herakleitos, Thales and Aristeia programmes co-financed by EU-ESF and the Greek NSRF, Greece;
BSF-NSF and MINERVA, Israel;
Norwegian Financial Mechanism 2014-2021, Norway;
NCN and NAWA, Poland;
La Caixa Banking Foundation, CERCA Programme Generalitat de Catalunya and PROMETEO and GenT Programmes Generalitat Valenciana, Spain;
G\"{o}ran Gustafssons Stiftelse, Sweden;
The Royal Society and Leverhulme Trust, United Kingdom.
 
The crucial computing support from all WLCG partners is acknowledged gratefully, in particular from CERN, the ATLAS Tier-1 facilities at TRIUMF/SFU (Canada), NDGF (Denmark, Norway, Sweden), CC-IN2P3 (France), KIT/GridKA (Germany), INFN-CNAF (Italy), NL-T1 (Netherlands), PIC (Spain), RAL (UK) and BNL (USA), the Tier-2 facilities worldwide and large non-WLCG resource providers. Major contributors of computing resources are listed in Ref.~\cite{ATL-SOFT-PUB-2023-001}.

% End of text imported from the .//acknowledgements/Acknowledgements.tex input file

% The next lines are included from the .//Acknowledgements-CMS.tex input file
CMS congratulates our colleagues in the CERN accelerator departments for the excellent performance of the LHC and thank the technical and administrative staffs at CERN and at other CMS institutes for their contributions to the success of the CMS effort. In addition, we gratefully acknowledge the computing centres and personnel of the Worldwide LHC Computing Grid and other centres for delivering so effectively the computing infrastructure essential to our analyses. Finally, we acknowledge the enduring support for the construction and operation of the LHC, the CMS detector, and the supporting computing infrastructure provided by the following funding agencies: SC (Armenia), BMBWF and FWF (Austria); FNRS and FWO (Belgium); CNPq, CAPES, FAPERJ, FAPERGS, and FAPESP (Brazil); MES and BNSF (Bulgaria); CERN; CAS, FRFCU, MoST, and NSFC (China); MINCIENCIAS (Colombia); MSES and CSF (Croatia); RIF (Cyprus); SENESCYT (Ecuador); MoER, ERC PUT and ERDF (Estonia); Academy of Finland, MEC, and HIP (Finland); CEA and CNRS/IN2P3 (France); BMBF, DFG, and HGF (Germany); GSRI (Greece); NKFIH (Hungary); DAE and DST (India); IPM (Iran); SFI (Ireland); INFN (Italy); MSIP and NRF (Republic of Korea); MES (Latvia); LAS (Lithuania); MOE and UM (Malaysia); BUAP, CINVESTAV, CONACYT, LNS, SEP, and UASLP-FAI (Mexico); MOS (Montenegro); MBIE (New Zealand); PAEC (Pakistan); MES and NSC (Poland); FCT (Portugal); MESTD (Serbia); MCIN/AEI and PCTI (Spain); MOSTR (Sri Lanka); Swiss Funding Agencies (Switzerland); MST (Taipei); MHESI and NSTDA (Thailand); TUBITAK and TENMAK (Turkey); NASU (Ukraine); STFC (United Kingdom); DOE and NSF (USA).

% End of text imported from the .//Acknowledgements-CMS.tex input file

\clearpage
\printbibliography
 
\clearpage
% ATLAS Collaboration author list
% Reference date of HIGG-2022-22 is 2023-05-09
% Author list last updated on date 18-DEC-23
% Data extracted on 18-Dec-2023 for paper reference HIGG-2022-22
% at 9:51am
 
\begin{flushleft}
\hypersetup{urlcolor=black}
{\Large The ATLAS Collaboration}

\bigskip

\AtlasOrcid[0000-0002-6665-4934]{G.~Aad}$^\textrm{\scriptsize 102}$,
\AtlasOrcid[0000-0002-5888-2734]{B.~Abbott}$^\textrm{\scriptsize 120}$,
\AtlasOrcid[0000-0002-1002-1652]{K.~Abeling}$^\textrm{\scriptsize 55}$,
\AtlasOrcid[0000-0001-5763-2760]{N.J.~Abicht}$^\textrm{\scriptsize 49}$,
\AtlasOrcid[0000-0002-8496-9294]{S.H.~Abidi}$^\textrm{\scriptsize 29}$,
\AtlasOrcid[0000-0002-9987-2292]{A.~Aboulhorma}$^\textrm{\scriptsize 35e}$,
\AtlasOrcid[0000-0001-5329-6640]{H.~Abramowicz}$^\textrm{\scriptsize 151}$,
\AtlasOrcid[0000-0002-1599-2896]{H.~Abreu}$^\textrm{\scriptsize 150}$,
\AtlasOrcid[0000-0003-0403-3697]{Y.~Abulaiti}$^\textrm{\scriptsize 117}$,
\AtlasOrcid[0000-0002-8588-9157]{B.S.~Acharya}$^\textrm{\scriptsize 69a,69b,m}$
% ...
\bigskip
\\

% ...
$^{29}$Physics Department, Brookhaven National Laboratory, Upton NY; United States of America.\\
% ...
$^{35}$$^{(a)}$Facult\'e des Sciences Ain Chock, R\'eseau Universitaire de Physique des Hautes Energies - Universit\'e Hassan II, Casablanca;$^{(b)}$Facult\'{e} des Sciences, Universit\'{e} Ibn-Tofail, K\'{e}nitra;$^{(c)}$Facult\'e des Sciences Semlalia, Universit\'e Cadi Ayyad, LPHEA-Marrakech;$^{(d)}$LPMR, Facult\'e des Sciences, Universit\'e Mohamed Premier, Oujda;$^{(e)}$Facult\'e des sciences, Universit\'e Mohammed V, Rabat;$^{(f)}$Institute of Applied Physics, Mohammed VI Polytechnic University, Ben Guerir; Morocco.\\
% ...
$^{49}$Fakult\"{a}t Physik , Technische Universit{\"a}t Dortmund, Dortmund; Germany.\\
% ...
$^{55}$II. Physikalisches Institut, Georg-August-Universit\"{a}t G\"ottingen, G\"ottingen; Germany.\\
% ...
$^{69}$$^{(a)}$INFN Gruppo Collegato di Udine, Sezione di Trieste, Udine;$^{(b)}$ICTP, Trieste;$^{(c)}$Dipartimento Politecnico di Ingegneria e Architettura, Universit\`a di Udine, Udine; Italy.\\
% ...
$^{102}$CPPM, Aix-Marseille Universit\'e, CNRS/IN2P3, Marseille; France.\\
% ...
$^{117}$Department of Physics, New York University, New York NY; United States of America.\\
% ...
$^{120}$Homer L. Dodge Department of Physics and Astronomy, University of Oklahoma, Norman OK; United States of America.\\
% ...
$^{150}$Department of Physics, Technion, Israel Institute of Technology, Haifa; Israel.\\
$^{151}$Raymond and Beverly Sackler School of Physics and Astronomy, Tel Aviv University, Tel Aviv; Israel.\\
% ...
$^{a}$ Also Affiliated with an institute covered by a cooperation agreement with CERN.\\
$^{b}$ Also at An-Najah National University, Nablus; Palestine.\\
$^{c}$ Also at Borough of Manhattan Community College, City University of New York, New York NY; United States of America.\\
$^{d}$ Also at Center for High Energy Physics, Peking University; China.\\
$^{e}$ Also at Center for Interdisciplinary Research and Innovation (CIRI-AUTH), Thessaloniki; Greece.\\
$^{f}$ Also at Centro Studi e Ricerche Enrico Fermi; Italy.\\
% ...
$^{m}$ Also at Department of Physics, King's College London, London; United Kingdom.\\
% ...

\end{flushleft}

% Created with Glance <Atlas.Glance@cern.ch>

\clearpage

% The next lines are included from the .//HIG-23-002-public-authorlist.tex input file
{\Large The CMS Collaboration}
 
\bigskip
\newskip{\cmsinstskip} \cmsinstskip=0pt plus 4pt

\cmsinstitute{Yerevan Physics Institute, Yerevan, Armenia}{\tolerance=6000
A.~Hayrapetyan, A.~Tumasyan\cmsAuthorMark{1}\cmsorcid{0009-0000-0684-6742}
\par}
\cmsinstitute{Institut f\"{u}r Hochenergiephysik, Vienna, Austria}{\tolerance=6000
W.~Adam\cmsorcid{0000-0001-9099-4341}, J.W.~Andrejkovic, T.~Bergauer\cmsorcid{0000-0002-5786-0293}, S.~Chatterjee\cmsorcid{0000-0003-2660-0349}, K.~Damanakis\cmsorcid{0000-0001-5389-2872}, M.~Dragicevic\cmsorcid{0000-0003-1967-6783}, A.~Escalante~Del~Valle\cmsorcid{0000-0002-9702-6359}, P.S.~Hussain\cmsorcid{0000-0002-4825-5278}
% ...
\par}
\vskip\cmsinstskip
\dag:~Deceased\\
$^{1}$Also at Yerevan State University, Yerevan, Armenia\\
% ...
$^{29}$Also at Bergische University Wuppertal (BUW), Wuppertal, Germany\\
% ...
$^{35}$Also at Institute of Physics, University of Debrecen, Debrecen, Hungary\\
% ...
$^{49}$Also at Helwan University, Cairo, Egypt\\
% ...
$^{55}$Also at Universit\`{a} di Napoli 'Federico II', Napoli, Italy\\
% ...
$^{69}$Also at Near East University, Research Center of Experimental Health Science, Mersin, Turkey\\
% ...
$^{102}$Also at Northeastern University, Boston, Massachusetts, USA\\
% ...

% End of text imported from the .//HIG-23-002-public-authorlist.tex input file

\end{document}